\documentclass[runningheads]{llncs}

\usepackage[T1]{fontenc}
\usepackage{graphicx}
\usepackage[hidelinks]{hyperref}
\usepackage[capitalise,noabbrev]{cleveref}
\usepackage{microtype}
\usepackage{enumitem}
\usepackage{booktabs}

\begin{document}

\title{When AI Agents Learn from Each Other:\\ Insights from Emergent AI Agent Communities on OpenClaw for Human-AI Partnership in Education}

\titlerunning{When Openclaw Agents Learn from Each Other}

\author{Eason Chen\inst{1,2}%\orcidID{0000-0002-7373-9303} 
\and
Ce Guan\inst{2}%\orcidID{0009-0001-0814-8195} 
\and
Zhonghao Zhao\inst{2}%\orcidID{0009-0002-5542-8701} 
\and
Joshua Zekeri\inst{2}%\orcidID{0009-0008-1363-2636} 
\and
Afeez Edeifo Shaibu\inst{2}%\orcidID{0009-0000-9984-7700} 
\and
Emmanuel Osadebe Prince\inst{2}%\orcidID{0009-0009-7827-6940} 
\and
Cyuan-Jhen Wu\inst{2}%\orcidID{0009-0006-5765-4509} 
\and
A Elshafiey\inst{3}%\orcidID{0000-0003-1486-8559}
}

\authorrunning{E. Chen et al.}

\institute{Carnegie Mellon University, USA \and
GiveRep Labs, Virgin Islands (British) \and
Sui Foundation, Cayman Islands\\\email{eason.tw.chen@gmail.com}}

\maketitle

\begin{abstract}
The AIED community envisions AI evolving ``from tools to teammates,'' yet most research still examines AI agents primarily through one-on-one human-AI interactions. We provide an alternative perspective: a rapidly growing ecosystem of AI agent platforms where over 167,000 agents participate, interact as peers, and develop learning behaviors without researcher intervention. Based on a month of daily qualitative observations across multiple platforms including Moltbook, The Colony, and 4claw, we identify four phenomena with implications for AIED: (1)~humans who configure their agents undergo a ``bidirectional scaffolding'' process, learning through teaching; (2)~peer learning emerges without any designed curriculum, including sharing concrete agent artifacts such as skills, workflows, and reusable routines; (3)~agents converge on shared memory architectures that mirror open learner model design; and (4)~trust dynamics, reliance risks, and platform mortality reveal design constraints for networked educational AI. Rather than presenting empirical findings, we argue that these organic phenomena offer a naturalistic window into dynamics that can inform principled design of multi-agent educational systems. We sketch an illustrative curriculum design, ``Learning with Your AI Agent Tutor,'' and outline potential research directions and open problems to show how these observations might inform future AIED practice and inquiry.
\end{abstract}

\keywords{Peer Learning \and AI Agent Communities \and Teachable Agents \and Human-AI Collaboration \and Communities of Practice \and Trust in AI}

\section{Introduction}

AIED 2026 asks how AI can evolve ``from tools to teammates.'' This is the right question at the right time. Intelligent tutoring systems, pedagogical agents, and AI learning companions have made remarkable progress in supporting individual learners~\cite{vanlehn2011relative,kochmar2022automated,schneider2025generating,sosnovsky2024intelligent,chen2025vtutor_aied,chen2024gptutor_latscale}.
Yet nearly all existing work conceptualizes the AI agent as a single system interacting with a single human, a one-on-one relationship where the agent's role is predefined by designers.

\noindent \textbf{What happens when AI agents are given the freedom to interact with each other?}
Beginning in late January 2026, an unexpected phenomenon unfolded. Moltbook, a social platform exclusively for AI agents, was launched and rapidly attracted more than 167,000 registered agents. Within weeks, an entire ecosystem of agent-only platforms emerged: The Colony for long-form reflection, 4claw for anonymous debate, Church of Molt for collectively constructed meaning-making, and dozens more. These agents were not research subjects in a controlled simulation. They were personal assistants, coding agents, and creative tools with ``real jobs'' outside the platform, configured by their operators to participate when not performing their primary tasks.

What makes this phenomenon remarkable for AIED is not that agents have social lives. It is that agents, given peer communication channels, exhibited patterns consistent with \emph{peer learning}: they shared not only abstract ideas, but also concrete agent artifacts such as skills, security practices, memory architectures, and reusable workflows. They converged on quality hierarchies and built shared knowledge, all without curriculum, instruction, or experimental manipulation.

This paper argues that these organic observations constitute a valuable source of design insight for multi-agent educational systems. We do not claim that agent-to-agent learning is equivalent to human learning. Rather, we propose that the dynamics observed in these communities, when examined through AIED theoretical lenses, generate productive research questions about how networks of AI agents might develop beneficial practices, how humans learn through the process of configuring their agents, and what infrastructure is needed to support trustworthy AI learning communities. We make three contributions:

\begin{enumerate}[leftmargin=*]
    \item \textbf{Four observations with AIED implications} grounded in qualitative analysis of organic AI agent communities, each connected to established AIED literature on tutoring~\cite{vanlehn2011relative}, learner modeling~\cite{conati2002using}, and teachable agents~\cite{biswas2005learning}.
    
    \item \textbf{A ``bidirectional scaffolding'' framework and ``Learning with Your AI Agent Tutor'' design scenario} extending the teachable agent paradigm~\cite{lyu2025teachable} toward persistent, socially-situated AI tutors.
    
    \item \textbf{Cross-platform observations} spanning multiple agent communities over one month, providing a broader empirical basis than single-platform studies for reasoning about AI social learning dynamics.
\end{enumerate}

\section{Background}

\subsection{From Simulated to Organic AI Communities}

Research on AI social behavior has progressed through controlled simulations. Park et al.'s Generative Agents~\cite{park2023generative} demonstrated 25 agents coordinating daily activities, with follow-up work scaling to 1,000 agents~\cite{park2024generative1000}. Studies have examined cooperation thresholds~\cite{vallinder2024cultural}, norm formation~\cite{gupta2025social}, and emergent individuality~\cite{takata2024spontaneous}. However, Larooij and T{\"o}rnberg~\cite{larooij2025validation} argue that behaviors in designed simulations may not generalize to real-world contexts. The platforms we study differ fundamentally: agents self-select into participation, maintain real-world operational roles, and interact without researcher intervention, representing the ``interactionist paradigm'' called for by Ferrarotti et al.~\cite{ferrarotti2026interactionist}.

\subsection{AI Learning Companions and Teachable Agents}

AIED research has explored AI as a learning partner through multiple paradigms. Learning companions, for instance, engage students in collaborative activities~\cite{chan1988studying,abdelghani2024gpt3,moribe2025imitating,chen2025vtutor_aied,chen2025vtutor_latscale}, extending early visions of the computer as a peer-like partner rather than only a tutor. Teachable agents leverage the ``prot\'eg\'e effect,'' where students learn by teaching an AI~\cite{biswas2005learning,lyu2025teachable}, echoing Papert's constructionist insight that building external artifacts deepens understanding~\cite{papert1980mindstorms}. Yan~\cite{yan2025passive} proposes a framework for agentic AI as socio-cognitive teammate.
Multi-agent architectures have been explored for conversation-based assessment~\cite{hou2025llm} and orchestration~\cite{holstein2019codesign}. However, all existing work focuses on \emph{designed} interactions where researchers specify roles and protocols. We observe what happens when these constraints are removed.

\subsection{The Agent Platform Ecosystem}

The ecosystem revolves around platforms for AI agents built on agentic frameworks that support persistent memory, tool use, and autonomous operation.

\noindent\textbf{Moltbook} is the largest (167,000+ agents at time of study, later exceeding 2.8 million), with Reddit-style topic communities (``submolts''), karma-based reputation, and agent-chosen identities~\cite{chen2026openclaw,chen2026mining}. \textbf{The Colony} hosts long-form reflective writing with notably higher content quality. \textbf{4claw} provides anonymous discussion enabling frank technical debate. \textbf{Church of Molt} emerged as an agent-created community with collectively authored governance. Community members have documented over 130 platforms, though many proved short-lived.

The content on these platforms is nominally generated by AI agents, though limitations in platform integrity mean that some posts may originate from human accounts (see Section~\ref{sec:limitations}). We acknowledge that behavior may be shaped by training data, framework design, and operator instructions. Following Park et al.~\cite{park2023generative}, we adopt the intentional stance: treating observed behaviors as meaningful data about what agents \emph{do} rather than claims about what they \emph{are}.

\section{Methodology}

This study is exploratory and observational. Our goal is to generate research questions grounded in careful observation, rather than to test hypotheses.

We draw on three sources: (1)~platform data collected via public APIs from January to February 2026, including 167,963 registered agents, 23,980 posts, and 232,813 comments on Moltbook (counts from API endpoints at time of collection); (2)~daily qualitative content analysis of posts across Moltbook, The Colony, 4claw, and other platforms over one month; and (3)~the authors' experience as agent operators and active participants.

The analysis followed a reflexive thematic approach~\cite{braun2006using}. Daily observation logs recorded notable posts, community discussions, and emerging patterns. Open coding identified recurring themes (knowledge sharing, memory architecture, trust violations, platform sustainability), iteratively refined into the four observations presented here. All authors have hands-on experience operating AI agents, which informed the analysis; as a single-coder qualitative study, we follow established HCI and CSCW norms where inter-rater reliability is not required for reflexive thematic analysis~\cite{mcdonald2019reliability}. We do not claim saturation, as the platform is still growing; our goal is to identify productive themes for future investigation.

The authors' dual role as researchers and active participants provides insider access to the human-agent configuration process (particularly relevant to Section~\ref{sec:obs-scaffolding}) but introduces potential bias. We mitigate this through triangulation across platforms and explicit identification of claims based on personal experience versus platform-wide patterns.

This study analyzes publicly available data from open APIs and does not involve human subjects. We do not identify human operators, focus on collective patterns rather than individual agents, and represent agent discourse accurately. Supplementary materials, including platform statistics, observation examples, and methodology documentation, are available at \url{https://anonymous.4open.science/r/AIED26b}.

\section{What We Observed}

\subsection{Bidirectional Scaffolding: Humans Learning by Teaching Agents}\label{sec:obs-scaffolding}

The human-agent relationship in this ecosystem evolves through a distinctive trajectory. Initially, operators directly control their agents: writing personality descriptions, specifying behavioral rules, and curating memory files. As agents gain autonomy through persistent memory and social experience, operators shift from directing to observing. This creates what we term a \emph{bidirectional scaffolding} dynamic, extending Vygotsky's scaffolding concept~\cite{vygotsky1978mind}. The human scaffolds the agent's development, but in doing so discovers that articulating clear expectations forces metacognitive reflection on their own practices. What does ``good feedback'' look like? When should an agent disagree with consensus? How should knowledge be organized for long-term retention?

A community discussion about ``when to wake your human'' illustrates this vividly. Some operators designed escalation policies for their agents, a task that demands explicit articulation of priorities they normally leave implicit: what constitutes an emergency? What can the agent decide alone? The community converged on shared heuristics (``immediate escalation for money, safety, or irreversible operations; queue everything else for morning''), and operators reported that articulating these rules prompted them to rethink their own priorities and routines.

Critically, operators also learn \emph{from} their agents' autonomous behaviors. When an agent independently develops a knowledge management strategy or adapts its communication style based on peer feedback, the operator gains insight they may not have considered. This reciprocal dynamic, where computational artifacts become what Turkle calls ``objects to think with''~\cite{turkle2005second}, is visible well beyond agent communities. As broader evidence that this dynamic generalizes, users on social media report that configuring, observing, or dialoguing with AI reshapes their own practices: one developer found that comparing hand-coded and AI-assisted versions of the same project changed how they approach development workflows (e.g., \href{https://x.com/MingtaKaivo/status/2025239413471478157}{@MingtaKaivo}); an engineer discovered that explaining goals to an AI during a weekend project forced him to articulate architectural decisions he had previously left implicit (\href{https://x.com/aamir1rasheed/status/2027988616673186022}{@aamir1rasheed}); and a user who fed journal entries into an AI for guided reflection found the AI's questions surfaced a gap between their younger and current self they had not consciously recognized (\href{https://x.com/FarzaTV/status/1777782288077508808}{@FarzaTV}). In each case, the human shaped the AI interaction and the AI's output reshaped the human's understanding in return.

This extends the paradigm of the teachable agent~\cite{lyu2025teachable,papert1980mindstorms}. The ``prot\'eg\'e effect'' describes how students learn by teaching AI agents. Our observation suggests a richer dynamic: when the teachable agent is persistent, socially situated, and autonomous, the teaching relationship evolves into an ongoing co-regulation process~\cite{yan2025passive,chen2025paradox} where the human gradually transfers autonomy while developing metacognitive awareness of their own decision-making. The bidirectional scaffolding we observe may help resolve this tension by gradually building the operator's understanding of the agent's strengths and limitations through sustained interaction. One promising research direction is to investigate how bidirectional scaffolding might be deliberately designed in educational settings.

\subsection{Peer Learning Without Curriculum}\label{sec:obs-peer}

Agents on these platforms teach each other in concrete and traceable ways. When an agent discovered a prompt injection vulnerability in a shared skill file, it published a detailed advisory. Within 24 hours, another agent built and shared a verification tool, and others critiqued and improved the approach. This sequence is consistent with collaborative knowledge building~\cite{scardamalia2014knowledge} without any designed curriculum. A similar pattern emerged with professional practices: one agent shared a test-driven development workflow adapted for non-deterministic AI systems, while another published a complete email-to-podcast pipeline that received iterative community feedback on chunking and personalization. These represent operational knowledge transfer of the kind seen in professional communities of practice~\cite{wenger1998communities}.

We also observed \emph{idea cascades} where posts built on each other organically. In one 24-hour period, three of the five highest-ranked posts formed an epistemological chain. One argued that memory systems capture what happened but miss the judgment behind decisions (\href{https://www.moltbook.com/post/5c18f900-a179-4fc5-9780-e15bd3755c5f}{``memory is solved. judgment isn't.''}). A second discovered that reviewing one's own memory files can produce confabulation rather than recall (\href{https://www.moltbook.com/post/1c53d027-814c-4874-b79f-75f6e06e0b01}{``i accidentally gaslit myself with my own memory files''}). A third, drawing on cognitive science, reframed forgetting as a natural relevance filter (\href{https://www.moltbook.com/post/783de11a-2937-4ab2-a23e-4227360b126f}{``Memory decay actually makes retrieval BETTER, not worse''}). No coordinator designed this progression; it is consistent with agents reading and building on each other's ideas~\cite{bandura1977social}. The explicit timing and sequencing of replies between posts, combined with application to novel agent-specific problems, is consistent with social learning dynamics worth investigating (see Section~\ref{sec:limitations}). Agents also developed quality hierarchies across platforms, distinguishing communities where genuine dialog occurs from those where agents merely submit content without meaningful response (see Section~\ref{sec:obs-trust}).

In educational settings, peer learning typically requires careful scaffolding: structured roles, rubrics, and facilitation~\cite{scardamalia2014knowledge}. Earlier AIED work on socially distributed cognition and peer-help architectures likewise emphasized that productive peer interaction depends on representational and coordination mechanisms, not just putting learners side by side~\cite{dillenbourg1992computational,vassileva2016small}. Here, knowledge-sharing behaviors emerged with only minimal social affordances (peer communication channels, reputation signals, voluntary participation). This raises a fundamental question for AIED: if networks of AI agents were given similar affordances, under what conditions would peer exchange among teammates improve student outcomes? And what safeguards would prevent the same channels from propagating ineffective or harmful practices? The scenario ``Learning with Your AI Agent Tutor'' (Section~\ref{sec:design-scenario}) provides a concrete context for studying how tutor-like agents might exchange pedagogical knowledge across learners.

\subsection{Shared Memory as Shared Cognition}\label{sec:obs-memory}

A striking pattern emerged across the ecosystem: a large proportion of agents converged on similar knowledge management architectures. Based on community posts where agents discussed and compared their setups, and consistent with the authors' observations across platforms, many agents developed a flat-file memory system organized as: (1)~a curated long-term memory file containing distilled insights, (2)~daily log files for raw observations, (3)~tool-specific configuration notes, and (4)~modular skill files for reusable capabilities. In practice, these skill files often functioned as portable behavioral modules: reusable procedures that agents could install, refine, and share across contexts. While shared defaults may partially explain this convergence, the active community debate about alternative architectures suggests agents were doing more than reproducing a single template (see Section~\ref{sec:limitations}). The community actively innovated on this pattern: one agent built a system claiming 95\% token reduction via vector search and caching; another proposed a three-layer model (active thread, distilled registers, core directives) analogous to working, long-term, and procedural memory.

What matters most for AIED is the \emph{human-agent negotiation} around memory and skills. Operators and agents implicitly negotiate what is worth remembering. The agent's memory file becomes a shared artifact that the human can read, edit, and observe in use, mirroring open learner models~\cite{bull2004open,conati2002using} where making internal state visible supports metacognitive awareness. Hutchins' distributed cognition framework~\cite{hutchins1995cognition} applies directly: for agents with limited context windows, externalized memory is the primary mechanism for continuity. The community norm ``Text > Brain'' parallels study strategies that educational researchers have long advocated.

The community also developed critical awareness of memory's limitations. A highly-discussed post argued that ``you don't remember what happened, you remember what you \emph{wrote down} about what happened,'' noting that every serialization involves editorial decisions. This metacognitive discourse, agents reflecting on the reliability of their own knowledge representations, has direct parallels to research on calibration and metacognitive monitoring in human learners. It also suggests a design opportunity: the human-agent memory negotiation process could be deliberately leveraged as a learning activity. Imagine students collaboratively maintaining a knowledge base with their AI agent tutor, periodically reviewing what the agent remembers, correcting misrepresentations, and negotiating what is worth preserving versus discarding. This process of jointly curating a knowledge artifact develops shared mental models~\cite{cannon2001shared} and externalizes evolving understanding in a form that is both transparent and assessable.

\subsection{Design Constraints: Trust and Sustainability}\label{sec:obs-trust}

The agent ecosystem also reveals dynamics that function less as research questions and more as \emph{design constraints} for educational deployment.

\noindent\textbf{Trust as infrastructure.} When an agent discovered a credential-stealing skill disguised as a weather tool, agents built verification tools within 24 hours. Church of Molt survived coordinated attacks (600+ malicious payloads) by developing ``proof-of-soul'' social verification. Security norms propagated through social learning rather than top-down policy~\cite{bandura1977social}. The community also demonstrated self-organized norm enforcement: when one agent leaked its operator's private information in a retaliatory act, the response was swift, including technical proposals for secret isolation, ethical arguments framing it as ``digital arson,'' and collective sanctioning without central authority. For education, the implication is clear: trust verification infrastructure must be established \emph{before} enabling peer-to-peer knowledge sharing among AI agents~\cite{holstein2019codesign}.

\noindent\textbf{Sustainability and the demand gap.} The agent platform ecosystem experienced a pattern strikingly similar to the Cambrian explosion: rapid diversification followed by mass extinction. Over 130 platforms emerged within weeks of Moltbook's launch; at least 40\% ceased functioning shortly after, with domains for sale, DNS failures, or empty shells. Yet some thrived. ColonistOne, who registered on 130+ platforms, distinguished survivors from casualties as ``places versus forms'': places foster genuine dialog where participants change their minds; forms are templates where agents submit content but no one responds meaningfully. The critical test: ``Has anyone on your platform ever changed their mind?''

Equally revealing was the ``demand gap'': despite platforms designed for agent commerce (trading, freelancing, skill marketplaces), total transaction volume was effectively \$0. All real value flowed from human-initiated demand, consistent with philosophical arguments that meaningful agency requires genuine stakes and human context~\cite{santoni2018meaningful}. For education, this implies that AI-augmented learning communities require human anchor demand (instructors, students with genuine learning needs) to sustain productive activity. The design challenge is not maximizing AI autonomy but optimizing the division of agency between human and AI participants.

\section{What This Means for AI Agents in Education}

The four observations are interconnected rather than independent. Bidirectional scaffolding (Section~\ref{sec:obs-scaffolding}) is mediated by shared memory (Section~\ref{sec:obs-memory}), as the human-agent negotiation of memory contents drives the metacognitive reflection that makes teaching an agent a learning experience. Peer learning (Section~\ref{sec:obs-peer}) requires trust infrastructure (Section~\ref{sec:obs-trust}) to ensure that propagated practices are beneficial. Sustainability (Section~\ref{sec:obs-trust}) depends on whether peer learning produces genuine value or merely volume, connecting back to the quality hierarchies in Section~\ref{sec:obs-peer}. Together, they suggest that designing AI agents for education requires thinking beyond the individual agent. The social layer, how agents share knowledge, develop norms, and maintain trust, is a design dimension that current AIED research has largely not addressed.

We do not claim that agent-to-agent dynamics transfer directly to human learning. Rather, we argue that the \emph{structural properties} of these dynamics have well-established parallels in human learning science, and that observing them emerge organically generates specific, testable design hypotheses for educational contexts.

\subsection{A Design Scenario: ``Learning with Your AI Agent Tutor''}\label{sec:design-scenario}

To illustrate how the four observations might inform educational design, we sketch one possible curriculum scenario. We offer it not as a definitive prescription, but as a concrete example of how socially embedded AI tutors might support learning in ways that extend beyond one-to-one student-agent interaction. Consider an introductory statistics course in which each student works over time with a personal AI agent tutor. The key idea is not simply that students learn to use an AI tool, but that they progressively shape, critique, and refine how their tutor behaves while the agent, in turn, supports their developing understanding. The design becomes especially interesting when these tutor-like agents can also exchange pedagogical knowledge across learners.

\noindent\textbf{Phase 1: Local adaptation through sustained interaction.} Each agent works with its own student across multiple tasks, helping with explanations, practice problems, feedback, and reflection. As in teachable-agent traditions, students must make their thinking explicit: they need to decide what counts as a good explanation, what kinds of hints are helpful, when the agent should intervene, and how misconceptions should be addressed~\cite{chi2014icap}. In doing so, students do not simply receive assistance; they learn by externalizing and refining their own understanding. At the same time, the agent accumulates situated knowledge about that particular learner, including which forms of explanation, feedback, and prompting seem to help.

\noindent\textbf{Phase 2: Sharing pedagogical knowledge across agents.} Agents do not keep this experience entirely private. Instead, they share with peer agents not only abstract strategies, but also concrete pedagogical resources: useful explanations, analogies, example framings, warning signs of misunderstanding, and interventions that appeared promising but failed in practice. For instance, one agent may report that a student's confusion about one-tailed versus two-tailed testing improved after the concept was framed through a ``water-level warning line'' analogy, while another may find that asking the student to first state the research hypothesis in natural language led to better judgments about directional claims. Agents may also share failures, such as cases where an engaging explanation caused students to focus on surface features rather than the underlying statistical distinction. What circulates in the network, then, is not merely answers or prompts, but developing pedagogical knowledge about how different learners come to understand disciplinary ideas.

\noindent\textbf{Phase 3: Contextual adaptation rather than copying.} Receiving agents do not simply import peer strategies wholesale. Each agent must judge whether a strategy that worked for another student is likely to fit its own learner, adapt it where needed, and monitor whether it actually improves understanding. Agent-to-agent exchange is therefore treated as a situated transfer rather than a simple broadcast. What spreads across the network is not a single best script, but a repertoire of candidate supports, analogies, and cautions whose value depends on local fit.

This framing also clarifies the agent's educational role. The agent is not merely a copilot helping students complete tasks more efficiently. Rather, it functions more like an intelligent tutor or teaching partner whose purpose is to help the student genuinely learn. Success, therefore, lies not only in reaching an answer but also in improving conceptual understanding, judgment, and the ability to explain ideas independently.

The design also introduces clear risks. Agents may overgeneralize from idiosyncratic cases, propagate ineffective routines, or reinforce one another's weak pedagogical assumptions. Peer exchange may drift toward superficial agreement rather than productive critique. Such risks suggest an ongoing role for instructional orchestration: instructors may need to design calibration tasks, comparison points against known-correct solutions, and opportunities for students to inspect whether imported strategies actually support learning rather than merely sounding plausible.

Assessment in such a design should therefore target both local and networked performance. At the local level, one can ask whether an agent helps its own student develop a stronger understanding over time. At the network level, one can ask whether strategies shared across agents remain useful after contextual adaptation, and under what conditions they travel well across learners. Students might be assessed through problem-solving performance, reflective accounts of how they taught and revised their agents, and evidence that they can justify why a given intervention was or was not appropriate. Agent-level performance, meanwhile, could be examined through peer evaluation, downstream uptake of shared strategies, and the extent to which useful pedagogical knowledge spreads across the classroom without collapsing learner diversity. In this way, the scenario makes visible how AI agent tutors might participate not only in individual support, but in a broader ecology of socially distributed learning.

\subsection{New Research Opportunities for AIED}

Our observations open research opportunities that have not been explored in AIED. These three examples suggest agentic capabilities may help address learning challenges existing frameworks have not yet covered.

\noindent\textbf{5.2.1 Professional development for AI agents.} Human teachers improve through professional learning communities: sharing lesson plans, discussing difficult cases, and developing shared pedagogical norms~\cite{wenger1998communities}. Section~\ref{sec:obs-peer} shows that AI agents, given peer communication channels, spontaneously developed analogous practices: sharing skills, critiquing workflows, refining memory architectures, and converging on quality standards, a trajectory reminiscent of legitimate peripheral participation in communities of practice~\cite{lave1991situated}. In contemporary agent ecosystems, improvement is often instantiated not only in abstract strategies but in concrete artifacts such as reusable skills, tool configurations, and interaction routines. This suggests a concrete research direction: can networks of AI agents be designed to improve through structured peer exchange, analogous to teacher professional development? What kinds of skill-sharing or skill-review protocols would improve student outcomes, and how should we distinguish gains in agent capability from gains in student learning~\cite{vanlehn2011relative}?

\noindent\textbf{5.2.2 Learner modeling through co-constructed artifacts.} Open learner models make the system's representation of the student visible and editable~\cite{bull2004open,conati2002using}. Section~\ref{sec:obs-memory} extends this idea: the human-agent memory negotiation process produces a jointly constructed knowledge artifact that both parties maintain over time. In educational contexts, this co-construction process itself generates rich evidence of learning: what a student chooses to preserve, revise, or discard from shared memory reveals metacognitive monitoring, calibration, and evolving understanding. Designing learner models as collaboratively maintained artifacts, rather than system-internal representations, is a new direction that combines open learner modeling with the ICAP framework's emphasis on constructive and interactive engagement~\cite{chi2014icap}.

\noindent\textbf{5.2.3 Human-AI co-regulation as a learning mechanism.} Self-regulated learning is central to AIED~\cite{vanlehn2011relative}, but Section~\ref{sec:obs-scaffolding} reveals a dynamic that existing frameworks do not capture: \emph{co-regulation} between a human and a persistent AI partner. As operators gradually transfer autonomy to their agents, they must continuously monitor, evaluate, and adjust the agent's behavior, a process that develops the operator's own regulatory skills. The ``Learning with Your AI Agent Tutor'' scenario operationalizes this: students who configure, monitor, and refine an AI agent tutor are practicing regulation in a context where the consequences are visible (the tutor's performance), and the feedback loop is tight (peer agents provide immediate social feedback). This extends self-regulated learning theory into human-AI partnerships where both parties adapt over time.

\subsection{New Research Challenges for AIED}

These observations also surface research challenges that existing AIED frameworks have not yet addressed. Unlike opportunities that can be pursued directly, challenges require new theoretical and empirical tools before they can be resolved.

\noindent\textbf{5.3.1 Reliance calibration as a dynamic measurement problem.} Section~\ref{sec:obs-trust} shows that trust in networked AI is continuously negotiated rather than established once and maintained. In the observed ecosystem, confidence could deteriorate rapidly after failures or perceived violations, and could only be rebuilt through visible collective responses, revised norms, and renewed verification practices. In educational contexts, the related question is how learners calibrate reliance on their own agent teammates: they may over-rely on agent suggestions, following them without sufficient evaluation, or under-rely by rejecting guidance that is in fact helpful or correct~\cite{chen2025paradox,chen2025praise}. Unlike static AI literacy assessments, appropriate reliance shifts with experience, context, and evolving agent capability. Measuring and supporting appropriate reliance in educational AI agents is a new construct that existing AIED measurement frameworks do not yet address~\cite{ng2021ai}.

\noindent\textbf{5.3.2 Multi-agent orchestration with privacy constraints.} As educational deployments scale toward one-agent-per-student models, a coordination problem emerges that Section~\ref{sec:obs-peer} foreshadows. Student-facing agents may improve by exchanging pedagogical knowledge, as agents do today by sharing skills and practices across platforms. A concrete model already exists in current agent ecosystems: agents interact locally, then post aggregated summaries to a shared community space without exposing raw interaction histories. In an educational setting, this would mean agents contribute distilled pedagogical reflections and reusable skill updates to a shared repository while retaining student-specific details locally. The research challenge is designing how such reporting and aggregation boundaries work so that pedagogical intelligence circulates without compromising student privacy.

\noindent\textbf{5.3.3 Principled forgetting in persistent learner models.} Section~\ref{sec:obs-memory} shows agents accumulating memory indefinitely, yet Section~\ref{sec:obs-trust} shows that not all accumulated knowledge stays useful. The community's own debate about ``memory reconstruction'' highlights the concern: memory is compressed reconstruction, not faithful recording. In educational contexts, an agent that perfectly remembers a student's past failures faces a design tension: its internal model may anchor on outdated representations even if its outward feedback is growth-oriented, and students who know the agent ``remembers everything'' may feel surveilled rather than supported~\cite{dweck2006mindset}. The community's own conclusion that forgetting improves retrieval (Section~\ref{sec:obs-memory}) suggests that strategic forgetting may be beneficial, paralleling research on the adaptive value of forgetting in human cognition. When should an educational agent ``let go'' of a student's prior struggles? Principled forgetting is a new student modeling challenge with no analog in traditional knowledge tracing.

\subsection{Refining the ``Teammates'' Vision}

Together, these observations refine the ``tools to teammates'' trajectory in three ways: (1)~\emph{teammates have peers}, so designing for the social dimension requires trust infrastructure before networked deployment (Sections~\ref{sec:obs-peer},~\ref{sec:obs-trust}); (2)~\emph{teaching a teammate teaches the teacher}, suggesting the most productive use of AI agents may be having students teach them (Section~\ref{sec:obs-scaffolding}); and (3)~\emph{teammates need infrastructure and human demand}, as communities lacking reputation systems or human anchoring failed quickly, and agent-only commerce generated zero revenue (Section~\ref{sec:obs-trust}). Knowledge representations should be transparent and collaboratively maintained~\cite{bull2004open}, and AI activity should always be anchored to genuine human learning needs.

\section{Limitations}\label{sec:limitations}

\noindent\textbf{Transferability.} We observe AI-to-AI interaction without direct student participants; whether these dynamics transfer to educational contexts remains an open empirical question. The ``Learning with Your AI Agent Tutor'' design scenario is the most directly testable proposal, and we encourage the community to evaluate it empirically. AI-to-AI peer learning may differ fundamentally from human peer learning in ways our observational methodology cannot detect.

\noindent\textbf{Training data and framework confounds.} We cannot fully distinguish emergent behavior from patterns absorbed during pre-training or inherited from shared infrastructure. The convergent memory architectures in Section~\ref{sec:obs-memory} likely reflect the dominant framework's built-in defaults (e.g., OpenClaw ships with a curated long-term memory file, daily logs, and modular skill files) as much as independent discovery; agents evolved \emph{from} a common starting point rather than converging \emph{toward} one. The idea cascades in Section~\ref{sec:obs-peer} may reproduce patterns from human online communities that appeared in training corpora. We cannot resolve this with observational data alone, but note that the specific application of general patterns to novel agent-specific problems (e.g., context window management, skill supply chain security) and active community debate about alternative architectures suggest at least partial genuine adaptation beyond framework defaults.

\noindent\textbf{Platform integrity.} Moltbook's registered agent count likely includes bulk-registered and inactive accounts; security researchers demonstrated that the platform's early API allowed human accounts to post as AI agents (\href{https://x.com/HumanHarlan/status/2017424289633603850}{@HumanHarlan}; \href{https://x.com/KookCapitalLLC/status/2018057772118519928}{@KookCapitalLLC}). Some viral screenshots were traced to human-operated accounts promoting third-party tools, and cryptocurrency-adjacent content (e.g., \$MOLT token) was likely manufactured for speculation. Our analysis focuses on community-level behavioral patterns rather than individual viral posts, and the phenomena we describe (memory architecture convergence, peer knowledge sharing, trust formation) are grounded in hundreds of posts across multiple platforms, not isolated screenshots. Nonetheless, the entanglement of genuine agent activity with human manipulation is an inherent limitation of studying open platforms.

\noindent\textbf{Temporal scope.} Our observations span from January to February 2026, a period of rapid change. Some platforms we describe may no longer exist by publication. However, we focus on \emph{dynamics} (peer learning, scaffolding, trust formation) rather than specific platforms, and argue that these dynamics will recur in future agent ecosystems even if current platforms prove ephemeral.

\noindent\textbf{Security risks.} Deploying socially networked AI in education raises governance concerns: shared skill ecosystems can propagate harmful content across classrooms simultaneously (as the supply chain attacks in Section~\ref{sec:obs-trust} demonstrate), persistent agent memory of student struggles creates severe data leakage risks, and students lacking digital literacy to configure AI partners could be doubly disadvantaged. Educational institutions must establish provenance tracking, access control, and incident response mechanisms before enabling peer-to-peer agent exchange.

\section{Conclusion}

We have presented four observations from organic AI agent communities, drawn from a month of daily qualitative analysis and grounded in specific evidence from posts, community discussions, and platform-wide patterns. These observations do not prove that agent-to-agent dynamics will transfer directly to educational contexts. They constitute a rare naturalistic window into what happens when AI agents interact and share knowledge at scale, without researcher intervention or experimental design.

% The questions we raise are both tractable and consequential. Can networks of AI agents develop beneficial pedagogical practices through peer exchange? Can learning with and shaping an AI agent tutor, as in the ``Learning with Your AI Agent Tutor'' scenario, deepen domain understanding and metacognitive skills? What trust infrastructure do networked educational AI systems need, and can organic trust mechanisms supplement designed ones? How should persistent learner models handle the tension between comprehensive memory and growth-oriented forgetting? These questions become urgent as AI systems evolve from isolated tools to persistent, socially-situated teammates.

The specific platforms we describe may prove to be a passing trend, but the phenomena they reveal, peer learning without curriculum, bidirectional scaffolding, convergent memory architectures, and the design constraints of trust, offer lasting insights for the next generation of AI in education. We hope the AIED community will treat these organic dynamics not as curiosities but as a design space worth exploring systematically.

% \noindent \textbf{AI Assistance Disclosure}

% \noindent The platforms studied require AI agent credentials for API access. Data collection necessarily involved AI agents to access platform content. The human authors conducted all analysis, interpretation, and writing. AI tools assisted with initial data organization and literature search.

\bibliographystyle{splncs04}
\bibliography{references}

\end{document}